# Nuclear shadowing in deep inelastic scattering on nuclei at very small x

Edgar V. Bugaev
*Institute for Nuclear Research of the Russian Academy of Sciences, Moscow 117312, Russia*

Boris V. Mangazeev
*Department of Physics, Irkutsk State University, Irkutsk 664003, Russia*

Nuclear shadowing corrections to the structure functions of deep inelastic scattering on intermediate-mass nuclei are calculated at very small values of Bjorken $x$ and small values of $Q^2$ $\left(Q^2 \leq 5\,GeV^2\right)$. The two-component approach developed in previous works of authors for a description of the nucleon structure functions of deep inelastic scattering is used. It is shown that the hard component of the nucleon structure functions that arises, in terms of the colour dipole model, from $q\bar{q}$ – pairs with a high transverse momentum, is almost not shadowed. It is shown that a change of the slope of the shadowing curve with a decrease of $x$ depends, at small values of $x$, on the relative contribution of the hard component to the nucleon structure function (this contribution is a function of $x$ and $Q^2$) and on a size of gluon saturation effects. It is shown that an accounting for saturation effects becomes essential for predictions of shadowing at $x < \left(10^{-5} - 10^{-4}\right)$, depending on a value of $Q^2$. Results of numerical calculations of nuclear shadowing for several nuclei are compared with available data of the E665 and the NMC collaborations.

## I. INTRODUCTION

It is well known that at small values of Bjorken $x$ ($x \leq 0.1$ for $Q^2 \geq 1\,GeV^2$) the inclusive structure function of deep inelastic scattering (DIS) on a nucleus with $A$ nucleons is smaller than the incoherent sum of the nucleon structure functions, i.e., $F_{2A} < AF_{2N}$. Correspondingly, the virtual photon-nucleus cross section is smaller than $A$ times the photon-nucleon cross section, $\sigma_{\gamma^*A} < A\sigma_{\gamma^*N}$. This, by definition, is the effect of nuclear shadowing (see reviews [1-5]). This phenomenon is explained by a destructive interference of amplitudes of single and multiple scatterings of the hadronic fluctuations of the virtual photon on nucleons of the target nucleus (considering the process in the rest frame of the nucleus). So, nuclear shadowing is a coherent effect and results from a coherent scattering of the hadronic fluctuation from at least two nucleons in the target nucleus.

There are two main formalisms which are used for a study of nuclear effects in DIS: Glauber-Gribov formalism [6,7] and Regge-Gribov framework [8-10]. In the first case the hadronic components of the virtual photon are rescattered in the target nucleus in a Glauber-like manner, and different models are distinguished by a choice of the mass spectrum of the hadronic fluctuations and a cross section of their interaction with nucleons. The Glauber-Gribov formalism had been exploited in calculations of nuclear shadowing based on the generalized vector dominance (GVD) approach [7,11-13] (the corresponding results of shadowing calculations are reviewed in [2]) and, more recently, in calculations using the colour dipole model [14,15] (for references on these calculations see reviews [3-5]).

Regge-Gribov framework uses the connection between nuclear shadowing and a differential cross section for the diffractive dissociation of the projectile. In order to calculate nuclear shadowing effects in this framework one must know the nucleon diffractive structure functions. Calculations of nuclear shadowing effects in Regge-Gribov framework are performed using the different model assumptions because the perturbative QCD is not applicable to the full description of diffractive DIS (especially in the region of small $Q^2$). Three groups of models are most frequently used for calculations of nuclear shadowing: i) aligned jet models (AJMs) [16,17], ii) Regge-motivated models using the concept of partonic pomeron [18,19] and iii) "leading twist approaches" operating with diffractive parton distributions [20] (the corresponding works studying shadowing are reviewed in [3-5]). Assumptions used in i) and ii) are not in conflict with QCD as it may seem. In opposite, it was shown [21] that perturbative QCD models based upon two-gluon exchange can be extrapolated into the nonperturbative region and, performing such an extrapolation, authors of [21] really discovered a dominance of aligned-jet configurations in the diffractive structure function and an arising of a simple picture of the Pomeron structure function.

In the present work we calculate the nuclear shadowing effects for several nuclei, for a broad interval of $x$ and a limited interval of $Q^2$, $Q^2 \leq 5\,GeV^2$. The choice of just this region of $Q^2$, for shadowing studies, is determined by two reasons. Firstly, in recent few years the interest was revived to a study of photonuclear interactions of high energy leptons (i.e., lepton-nucleus inelastic interactions dominated by small values of $Q^2$). This interest is connected, in particular, with a planning of new experiments on detection of astrophysical neutrinos of very high energies (see, e.g., [22,23]). The second reason is the following: in our previous papers [24-26] we elaborated the two-component model (GVD + perturbative QCD) for a description of DIS just for a region of small and medium $Q^2$ and, in the present work, we use, for a calculation of the shadowing corrections, the nucleon structure functions of DIS obtained in this two-component model.

It is well known that shadowing effects in interactions of real and quasireal photons with nuclei are rather well described by the vector dominance approach operating with light vector mesons only ($\rho, \omega, \varphi$) (see, e.g., the detailed review [27]). However, when data with photons of relatively high virtualities ($Q^2 \geq 1\,GeV^2$) appeared, it had been realized that the light mesons explain well the "high twist" shadowing effects while for a description of a weak $Q^2$-dependence of shadowing at medium and large $Q^2$, discovered by the data, the additional "quasiscaling" or "partonic" mechanism [28] is necessary. The alternative (and more appropriate, in our opinion) approach for a description of shadowing at medium $Q^2$ is an application of the GVD concept in aligned jet version [16,17,29]. In GVD excited states of light vector mesons ($\rho', \rho'',...$) are included in the mass spectrum of the hadronic fluctuations of the virtual photon [11,13]. In such an approach the photoabsorption cross section $\sigma_{\gamma^* N}$ contains the GVD (i.e., non-perturbative) part and the perturbative term taking into account a contribution of those hadronic fluctuations of the virtual photon whose interactions with the target nucleon are described by perturbative QCD. Just this approach is used for calculations of the DIS structure functions in [24-26].

The plan of the paper is as follows. In the second Section the main assumptions underlying our approach are discussed and some key formulas are briefly derived. In the third Section the shadowing correction due to soft interactions of the hadronic fluctuations of the virtual photon with nucleons of the target nucleus is studied. In the fourth Section the contribution to shadowing from the hard interactions of non-aligned $q\bar{q}$-pairs, produced by the virtual photon, with nucleons is considered. In the fifth Section results of the numerical calculations of the shadowing coefficients for several nuclei and for a broad interval of Bjorken $x$ are shown. Discussions of the results and conclusions are given in the sixth Section.

## II. OUTLINE OF THE MODEL

Consider, at first, a simplest case when the hadronic fluctuations are described by the separate vector mesons.

GVD approach (see, e.g., [30] for the historical review) starts from the spectral representation for the transverse photon absorption cross section $\sigma_T(s, Q^2)$ ($s$ is the square of the virtual photon-nucleon center of mass energy),

$$\sigma_T(s, Q^2) = \int dM^2 \frac{M^4}{(M^2 + Q^2)^2} \rho_T(s, M^2) \quad (1)$$

(the longitudinal part of the photoabsorption cross section, $\sigma_L$, is considered below, in Sec.3).

The spectral weight function $\rho_T$ is related to the imaginary part of the forward scattering amplitude for $V(M) + N \rightarrow V(M) + N$ ($V$ is the vector meson state with mass M). The main relation of GVD is [7,11]

$$\rho_T(s, M^2) = \frac{1}{4\pi^2 \alpha_{em}} \sigma_{had}(M^2) \sigma_{VN}(s, M^2), \quad (2)$$

where

$$\sigma_{had}(M^2) = \sigma_{e^+ e^- \rightarrow hadrons}(M^2) \quad (3)$$

The vector meson-photon couplings are defined by the connection (in the approximation of zero decay width of the vector mesons)

$$\sum_n \frac{e^2}{f_n^2} \delta(M^2 - M_n^2) = \frac{1}{4\pi^2 \alpha_{em}} \sigma_{had}(M^2). \quad (4)$$

Here we assume, in accordance with GVD, that there is at least one family of neutral vector mesons (in reality, we consider radial excitations of $\rho$ only).

Introducing the ratio

$$R(M^2) = \frac{\sigma_{had}(M^2)}{\sigma_{e^+ e^- \rightarrow \mu^+ \mu^-}(M^2)} = \sum_n R_n(M^2), \quad (5)$$

one obtains from eq. (4):

$$R_n(M^2) = \frac{12\pi^2}{f_n^2} M_n^2 \delta(M^2 - M_n^2). \quad (6)$$

For taking into account the nonzero decay width one must replace $\delta$-function in eq.(6) by the Breit-Wigner-type expression:

$$\pi\delta\left(M^2 - M_n^2\right) = \frac{M_n \Gamma}{\left(M^2 - M_n^2\right)^2 + M_n^2 \Gamma^2}. \quad (6a)$$

We assume, further, in accord with QCD (in its large $N_c$ limit) and Regge theory (see, e.g., [31]), that the mass squared of the family member obeys "equal spacing rule" with respect to the index $n$,

$$M_n^2 = M_0^2 (1 + an), \quad n = 0, 1, 2... \quad (7)$$

With such a spectrum the parton-hadron duality condition,

$$R(M^2) \cong const, \quad (8)$$

leads to the following relation for the photon-vector meson couplings of the family members:

$$\frac{1}{f_n^2} = \frac{1}{f_\rho^2} \frac{M_\rho^2}{M_n^2}. \quad (9)$$

Note, that we do not need the analogous relation for the electromagnetic decay widths of the vector mesons (see the recent work [32] where the question of the validity of eq. (9) is discussed).

Substituting eq. (2) in eq. (1) we obtain the relation

$$\sigma_T(s, Q^2) \sim \sum_n \frac{1}{f_n^2} \frac{M_n^4}{\left(M_n^2 + Q^2\right)^2} \sigma_{V_n N}(s) \quad (10)$$

and, using the Glauber-Gribov formalism, calculate the cross section on a nucleus. In leading order (i.e., for the coherent scattering on two nucleons) one obtains the expression

$$\sigma_T^A = A\sigma_T - \pi A(A-1)$$
$$\times \sum_n \frac{\alpha_{em}}{f_n^2} \frac{M_n^4}{\left(M_n^2 + Q^2\right)^2} \sigma_{V_n N}^2 C(M_n), \quad (11)$$

where $C(M_n)$ is the nuclear factor depending on nucleon densities inside of the nucleus and on the coherent length (see, e.g., [33]),

$$C(M_n) = \int d^2b \left|\Phi(k_L, b)\right|^2, \quad (12)$$

$$\Phi(k_L, b) = \int dz \rho(b, z) e^{ik_L z}, \quad (13)$$

$$k_L = xm_N \left(1 + \frac{M_n^2}{Q^2}\right) = \frac{Q^2 + M_n^2}{2\nu}. \quad (14)$$

The same expression for $\sigma_T^A$ can be obtained using the Regge-Gribov framework. Here, one needs the differential cross section for the diffractive production which, in the GVD approach (assuming the diagonal approximation, as in eq. (1), and the approximation of a zero width, as in eq. (4)), is given by the formula [34]

$$\left.\frac{d^2\sigma_{\gamma^*N}^{diff}}{dM_X^2 dt}\right|_{t=0} = \sum_n \frac{\alpha}{4}\left(\frac{M_n}{f_n}\right)^2 \delta\left(M_X^2 - M_n^2\right)$$
$$\times \frac{M_X^2}{\left(M_X^2 + Q^2\right)^2} \sigma_{XN}^2. \quad (15)$$

Using this formula, the cross section $\sigma_T^A$ is easily calculated from the main expression of the Regge-Gribov framework [35],

$$\sigma_A^T = A\sigma_T - 4\pi A(A-1) \int dM_X^2 \left.\frac{d^2\sigma_{\gamma^*N}^{diff}}{dM_X^2 dt}\right|_{t=0} C(M_X). \quad (16)$$

In these equations, $M_X$ is the invariant mass of the $X$-system produced in the diffractive process.

As mentioned in the Introduction, we use in the present work the aligned jet version of GVD. In this version, all members of GVD sums, in particular, in eqs. (10) and (11), are multiplied on the cutting factors inversely proportional to $M_n^2$. It corresponds, in the colour dipole picture, to a reducing of the phase space of $q\bar{q}$-pairs produced by the virtual photon, and provides the approximate $Q$-scaling, $Q^2 \sigma_{\gamma^*N} \sim const$. $q\bar{q}$-pairs with a small transverse momentum have large transverse size and interact with the target nucleon with a large cross section, and, just by this reason, we may assume that $\sigma_{V_n N}$ in eq.(10) doesn't depend on the meson mass. The cutting factors depend on the ratio $\frac{k_{0\perp}^2}{M_n^2}$, and, in a leading order, are [26]:

$$\eta_T^n \simeq 3\frac{k_{0\perp}^2}{M_n^2}, \quad \eta_L^n \simeq 6\left(\frac{k_{0\perp}^2}{M_n^2}\right)^2, \quad (17)$$

a value of $k_{0\perp}$ is the model parameter, and it was taken equal to $0.385\,GeV$ in [26]. Note, that it is close to the value of the perturbative QCD scale $\Lambda_s$, $\Lambda_s = 0.339\,GeV$ [36]. For a transition to the aligned jet version one must do replacements

$$\frac{e^2}{f_n^2} \to \frac{e^2}{f_n^2} \eta_{T,L}^n \quad (18)$$

in GVD expressions for $\sigma_{T,L}$.

In calculations of the nucleon structure functions of DIS, in the region of small and medium $Q^2$, the number of those vector mesons which saturate the GVD sums is around 8-9, if their masses are given by eq.(7) with $a = 2$. We assume that those vector mesons which almost do not contribute to the total photoabsorption cross section (but contribute noticeably to the diffractive cross section) form a high-mass continuum. For a description of the diffractive cross section in the region of large invariant masses $M_X$ it is natural to use Regge parameterization. Concretely, the triple-pomeron limit should be good enough. The corresponding parameterization is rather simple,

$$s\frac{d^2\sigma_{\gamma^*N}^{diff}}{dM_X^2 dt}\bigg|_{t=0} \sim s^{2\alpha_P(0)-1}\left(\frac{1}{M_X^2+Q^2}\right)^{\alpha_P(0)}. \quad (19)$$

Here, $\alpha_P(0)$ is the soft pomeron intercept. Another way is to use the available parameterizations of the experimental data and extrapolate them to a region of large values of $s$ and $M_X^2$.

As stated above, an application of aligned jet version of GVD for a calculation of the nucleon structure function leads, in a natural way, to a two-component picture: the total $\sigma_{\gamma^*N}$, in particular, contains, except of the nonperturbative ("soft") component, the perturbative ("hard") component which arises from the contribution of $q\bar{q}$-pairs with a high transverse momentum.

So, now, for a case of the transverse virtual photons one has the sum:

$$\sigma_T = \sigma_T^{soft} + \sigma_T^{hard}, \quad (20)$$

where $\sigma_T^{soft}$ is given by eq.(10) (with replacements introduced in eq. (18)). From here, and everywhere below in the text, we change the notation designating by $\sigma_T$ just the sum of the soft and hard parts.

If we assume, on a moment, that hadronic configurations which constitute the hard component interact completely incoherently with nucleons of the target nucleus (due to the colour transparency phenomenon), then we have, in an approximation of the pure $\rho$-dominance, the simple expression [1] for the shadowing effect:

$$A_{eff} = \frac{\sigma_T^A(s,Q^2)}{\sigma_T(s,Q^2)} = \left[1-\lambda(s,Q^2)\right]$$
$$\times \frac{\sigma_{\rho A}(s)}{\sigma_{\rho N}(s)} + \lambda(s,Q^2)A. \quad (21)$$

where

$$\lambda = \frac{\sigma_T^{hard}(s,Q^2)}{\sigma_T(s,Q^2)}. \quad (22)$$

In a general case, however, one must take into account the shadowing effect from the hard component and, also, include into a consideration the contribution from longitudinal virtual photons. By definition, the shadowing coefficient is given by the general formula

$$\alpha = \frac{F_{2A}}{AF_{2N}} = \frac{\sigma_{\gamma^*A}}{A\sigma_{\gamma^*N}}, \quad (23)$$

where the nucleon structure function $F_{2N}$ and the nuclear structure function $F_{2A}$ are simply connected with the photoabsorption cross sections:

$$F_{2N} = \frac{Q^2}{4\pi^2\alpha_{em}}\sigma_{\gamma^*N}, \quad F_{2A} = \frac{Q^2}{4\pi^2\alpha_{em}}\sigma_{\gamma^*A}. \quad (24)$$

It is convenient, for more clarity, to introduce two shadowing coefficients, soft and hard ones,

$$\alpha_{soft} = \frac{F_{2A}^{soft}}{AF_{2N}^{soft}}. \quad (25)$$

$$F_{2A}^{soft} = F_{2A}^{(\rho,\rho'...)} + F_{2A}^{(cont)},$$
$$F_{2N}^{soft} = F_{2N}^{(\rho,\rho'...)}, \quad (25a)$$

and

$$\alpha_{hard} = \frac{F_{2A}^{hard}}{AF_{2N}^{hard}}. \quad (26)$$

If there is no shadowing in the hard part then $\alpha_{hard}=1$. The total shadowing coefficient is given by the sum:

$$\alpha = \alpha_{soft}\frac{F_{2N}^{soft}}{F_{2N}} + \alpha_{hard}\frac{F_{2N}^{hard}}{F_{2N}}. \quad (27)$$

### III. SHADOWING OF THE SOFT COMPONENT

According to the previous Section, the soft component of the nuclear structure function (and the corresponding photoabsorption cross section) contains the vector meson part and the high-mass continuum part,

$$\sigma_{\gamma^*A}^{soft} = \sigma_{\gamma^*A}^{(\rho,\rho'...)} + \sigma_{\gamma^*A}^{(cont)}. \quad (28)$$

Consider, at first, the vector meson part. The shadowing correction is given by the quantity

$$\delta\sigma_{\gamma^*A}^{(\rho,\rho'...)} = A\sigma_{\gamma^*N}^{(\rho,\rho'...)} - \sigma_{\gamma^*A}^{(\rho,\rho'...)} \quad (29)$$

The contribution to this correction from the transverse photons can be extracted from eq. (11):

$$\delta\sigma_{\gamma_T^*A}^{(\rho,\rho'...)} = \pi A(A-1)$$
$$\times \sum_n \frac{\alpha_{em}}{f_n^2}\frac{M_n^4}{(M_n^2+Q^2)^2}\sigma_{V_nN}^2 C(M_n). \quad (30)$$

For taking into account the higher rescattering terms we introduce in the right part of eq. (11) an eikonal factor $F$ (inside of the integral over the impact parameter in eq. (12)). This factor is

$$F = e^{-\frac{A}{2}\sigma_{eff}T(b)} \quad (31)$$

Here, $T(b)$ is the nuclear thickness,

$$T(b) = \int_{-\infty}^{\infty} dz \rho(b,z), \quad (32)$$

and $\sigma_{eff}$ is the effective cross section for the interaction of the diffractively produced state with the nucleon. In our case, evidently, $\sigma_{eff} = \sigma_{V_n N}^T \equiv \sigma_{V_n N}$. Besides, we insert in eq. (11) the cutting factors $\eta_T^n$. Finally, one obtains, using eq. (13), the expression

$$\delta\sigma_{\gamma_T^* A}^{(\rho,\rho'...)} = 2\pi A(A-1) \sum_n \frac{\alpha_{em}}{f_n^2} \eta_T^n \frac{M_n^4}{(Q^2 + M_n^2)^2} \sigma_{V_n N}^2$$

$$\times \int d^2 b \int_{-\infty}^{\infty} dz_1 \int_{z_1}^{\infty} dz_2 \rho(b,z_1) \rho(b,z_2) \cos(k_L(z_2 - z_1)) \quad (33)$$

$$\times \exp\left[-\frac{A}{2}\sigma_{V_n N} \int_{z_1}^{z_2} \rho(b,z)dz\right].$$

The longitudinal part of $\delta\sigma_{\gamma^* A}^{(\rho,\rho'...)}$ is derived analogously. It can be obtained from eq. (33) by substitutions

$$M_n^4 \to Q^2 M_n^2, \; \sigma_{V_n N} \to \xi \sigma_{V_n N}, \; \eta_T^n \to \eta_L^n. \quad (34)$$

Here, the factor $\xi$ is the ratio $\sigma_{V_n N}^L / \sigma_{V_n N}^T$ (see [26] for details).

The contribution of the high-mass continuum into the soft shadowing correction is given by the formula followed from eq. (16):

$$\delta\sigma_{\gamma^* A}^{(cont)} = 4\pi A(A-1) \int_{M_{X,\min}^2} dM_X^2 \frac{d^2\sigma_{\gamma^* N}^{diff}}{dM_X^2 dt}\bigg|_{t=0} C(M_X),$$

$$(35)$$

where the nuclear factor $C(M_X)$ defined above is calculated from eqs. (12-14) with a substitution $M_n \to M_X$. Here, for calculations of the diffractive cross section, we prefer to use instead the Regge triple-limit formula the available parametrizations of the direct ZEUS data [37,38]. Concretely, these parametrizations were done for the diffractive nucleon structure function $F_2^{D(3)}$ connected with the diffractive cross section by the relation

$$\frac{d^2\sigma_{\gamma^* N}^{diff}}{dM_X^2 dt}\bigg|_{t=0} \cong \frac{4\pi^2 \alpha_{em}}{Q^4} xB F_2^{D(3)}(x,Q^2,M_X^2), \quad (36)$$

Here, $B$ is the slope of the $t$-dependence of the diffractive cross section,

$$\frac{d^2\sigma_{\gamma^* N}^{diff}}{dM_X^2 dt} = \frac{d^2\sigma_{\gamma^* N}^{diff}}{dM_X^2 dt}\bigg|_{t=0} e^{Bt}. \quad (37)$$

For a comparison with experimental data it is more convenient to use the following variables:

$$x_P = \frac{M_X^2 + Q^2}{s + Q^2} \simeq \frac{M_X^2}{s}, \; \beta = \frac{x}{x_P} \simeq \frac{Q^2}{M_X^2}. \quad (38)$$

To our knowledge, the most recent parameterizations which are valid for a broad interval of values of the variables are based on the well-known BEKW model [21]. Specifically, (neglecting the small contribution from the longitudinal virtual photons) one has [37,38]

$$x_P F_2^{D(3)}(\beta, x, Q^2) \approx C_T F_{q\bar{q}}^T + C_g F_{q\bar{q}g}^T,$$

$$F_{q\bar{q}}^T = \left(\frac{x_0}{x_P}\right)^{n_T(Q^2)} \beta(1-\beta), \quad (39)$$

$$F_{q\bar{q}g}^T = \left(\frac{x_0}{x_P}\right)^{n_g(Q^2)} \ln\left(1 + \frac{Q^2}{Q_0^2}\right)(1-\beta)^\gamma.$$

Here

$$n_{T,g}(Q^2) = n_0 + n_1 \ln\left(1 + \frac{Q^2}{Q_0^2}\right), \quad (39a)$$

and $C_T, C_g, n_0, n_1, Q_0^2, x_0, \lambda$ are parameters the fit. Note, that in the limit $\beta \to 0$ this parameterization gives, approximately

$$s \frac{d^2\sigma_{\gamma^* N}^{diff}}{dM_X^2 dt}\bigg|_{t=0} \sim \left(\frac{1}{x_P}\right)^{n_0+1} \sim \left(\frac{1}{M_X^2 + Q^2}\right)^{n_0+1}, \quad (40)$$

This is close to the triple-limit Regge formula, eq. (19), if $n_0 \approx \alpha_P(0) - 1$.

Finally, the shadowing correction due to the soft interactions of the hadronic fluctuations of the virtual photons with the target nucleus is given by the formula

$$\alpha_{soft} = 1 - \frac{\delta\sigma_{\gamma^* A}^{(\rho,\rho'...)} + \delta\sigma_{\gamma^* A}^{(cont)}}{A\sigma_{\gamma^* N}^{soft}}, \quad (41)$$

where $\delta_{\gamma^* A}^{(\rho,\rho'...)}$ and $\delta_{\gamma^* A}^{(cont)}$ are calculated using eqs.(33,35).

### IV. SHADOWING OF THE HARD COMPONENT

The $q\bar{q}$-pairs, produced by the virtual photons, with high transverse momenta have relatively small transverse sizes and their interaction with the nucleon can be described, in a language of Regge theory, by an exchange of the perturbative ("hard") pomeron [39]. In terms of the colour dipole model, the photoabsorption cross sections on the nucleon are given by the integrals

$$\sigma_{T,L}(s,Q^2) = \int dz \int d^2 r_\perp \left|\psi^{T,L}(r_\perp, z, Q^2)\right|^2 \sigma_{q\bar{q}N}^{hard}(r_\perp, s). \quad (42)$$

Here, $\psi^{T,L}$ are the light-cone wave functions of the virtual photon, $\sigma_{q\bar{q}N}^{hard}$ is the total cross section of the hard interaction of the $q\bar{q}$-pair with the nucleon, $r_\perp$ is the transverse separation of particles of the pair, $z$ is the fraction of the incoming photon light cone energy for one quark of the pair. For a phenomenological description of the hard interaction part of $\sigma_{q\bar{q}N}^{hard}$ we use the FKS model [40], in which this cross section has the following functional form at $r_\perp \to 0$:

$$\sigma_{q\bar{q}N}^{hard} \sim r_\perp^2 e^{-ar_\perp}\left(r_\perp^2 s\right)^b \qquad (43)$$

and it is assumed that the $s$-dependence of this cross section is more strong than in a case of the soft pomeron.

The corresponding shadowing correction is calculated with a help of the general formula of eq. (16). In this formula, the nuclear factor $C\left(M_X^2\right)$ depends on $M_X$ only through $k_L$. We assume that $M_X^2$ is, approximately, equal to $Q^2$, due to a chain of approximate equalities:

$$M_X^2 \sim k_\perp^2 \sim r_\perp^{-2} \sim Q^2. \qquad (44)$$

If $M_X^2 = Q^2$, one has $k_\perp = 2xm_N$ and the factor $C\left(M_X^2\right)$ doesn't depend on $M_X$. Therefore, one can integrate the diffractive cross section in the integrand of eq. (16) over $M_X^2$ and use the known expression of the colour dipole model [41]:

$$\left.\frac{d^2\sigma_{\gamma_{L,T}^*N}^{diff}}{dt}\right|_{t=0} = \frac{1}{16\pi}\int dz \int d^2r_\perp \left|\psi^{L,T}\left(r_\perp,z,Q^2\right)\right|^2 \qquad (45)$$

$$\times \left[\sigma_{q\bar{q}N}^{hard}\left(r_\perp,s\right)\right]^2.$$

Finally, the shadowing correction for the hard component is given by the expression (summing over photon polarizations)

$$\delta\sigma_{\gamma^*A}^{hard} = \frac{A(A-1)}{2}\int dz \int d^2r_\perp \left|\psi\left(r_\perp,z,Q^2\right)\right|^2$$

$$\times \left[\sigma_{q\bar{q}N}^{hard}\left(r_\perp,s\right)\right]^2 \int d^2b \int_{-\infty}^{\infty} dz_1 \int_{z_1}^{\infty} dz_2 \rho(b,z_1)\rho(b,z_2) \qquad (46)$$

$$\times \cos\left(2xm_N(z_1-z_2)\right)e^{-\frac{A}{2}\sigma_{q\bar{q}N}^{hard}(r_\perp,s)\int_{z_1}^{z_2} dz\rho(b,z)}.$$

Here we used the notation:

$$\left|\psi\left(r_\perp,z,Q^2\right)\right|^2 \equiv \left|\psi^T\left(r_\perp,z,Q^2\right)\right|^2 + \left|\psi^L\left(r_\perp,z,Q^2\right)\right|^2. \qquad (47)$$

For the quark dipoles with small transverse size the formulas for the virtual photon wave function are derived from QED (see, e.g., [42]),

$$\left|\psi^T\left(r_\perp,z,Q^2\right)\right|^2 = \frac{3\alpha}{2\pi^2}\sum_f \left(\frac{e_f}{e}\right)^2 \left[m_f^2 K_0^2(\varepsilon r_\perp)\right. \qquad (48)$$

$$\left.+\left(z^2+(1-z)^2\right)\varepsilon^2 K_1^2(\varepsilon r_\perp)\right],$$

$$\left|\psi^L\left(r_\perp,z,Q^2\right)\right|^2 = \frac{6\alpha_{em}}{\pi^2}\sum_f \left(\frac{e_f}{e}\right)^2 \left(Q^2 z^2(1-z)^2\right) K_0^2(\varepsilon r_\perp) \qquad (49)$$

$$\varepsilon^2 = z(1-z)Q^2 + m_f^2. \qquad (50)$$

The hard shadowing coefficient defined in eq. (26) is given by the expression

$$\alpha_{hard} = 1 - \frac{\delta\sigma_{\gamma^*A}^{hard}}{A\sigma_{\gamma^*N}^{hard}}, \quad \sigma_{\gamma^*N}^{hard} = \sigma_T^{hard} + \sigma_L^{hard}, \qquad (51)$$

and $\sigma_{T,L}^{hard}$ are calculated using eq. (42).

## V. RESULTS OF CALCULATIONS OF SHADOWING COEFFICIENTS

According to derivations of the previous Sections the total shadowing coefficient is calculated using the expression

$$\alpha = 1 - \frac{\delta\sigma_{\gamma^*A}^{(\rho,\rho'...)} + \delta\sigma_{\gamma^*A}^{(cont)} + \delta\sigma_{\gamma^*A}^{hard}}{A\left(\sigma_{\gamma^*N}^{soft} + \sigma_{\gamma^*N}^{hard}\right)}, \qquad (52)$$

In our previous work [43] we supposed (in calculations of shadowing) that the soft hadronic fluctuations of the virtual photon consist of the one separate vector meson ($\rho_0$) and the continuum with the border mass 1.5 GeV. In the present work we used the GVD approach and, respectively, took into account excited states of the $\rho$-meson family (eight mesons, in addition to $\rho_0$) with masses determined by eq. (7) with $a$ =2. The border mass of the continuum is equal to 3.3 GeV.

Calculating shadowing corrections from separated vector mesons we should, for a comparison with data in the region of large values of Bjorken $x$, slightly modify eq. (33) and its analogue in a case of the longitudinal photon, inserting into their integrands the "damping factor" (see, e.g., [44,45]). This factor is necessary because in the region of large $Q^2$, $Q^2 \gg 1 GeV^2$, and large Bjorken $x$ ($x \sim 10^{-2}$) the vector dominance approach is too rough: in this region the coherent length is small while $q\bar{q}$-pairs from the virtual photon are too narrow ($r_\perp^2 \sim Q^{-2}$) and, therefore, the cross section of their interaction with the target nucleon does not have time to grow to a value of the order of $\sigma_{V_nN}$. We use the following form of this damping factor:

$$H(Q^2, x) = \frac{1}{1 + \frac{Q^2}{Q_0^2(x)}}, \quad Q_0^2(x) = 1.5\left(\frac{10^{-2}}{x}\right). \quad (53)$$

For a calculation of the shadowing correction due to the high mass continuum we use eqs. (35-39) with the following values of the parameters [38]:

$$C_T = 0.072; \ C_g = 0.008; \ n_0 = 0.13; \ n_1 = 0.053;$$
$$Q_0^2 = 0.4; \ x_0 = 0.01; \ \gamma = 12.78. \quad (54)$$

At last, for a calculation of shadowing of the hard component we use the corresponding formula of the FKS model [40]

$$\sigma_{q\bar{q}N}^{hard}(r_\perp^2, s) = \left(\alpha_2^H r_\perp^2 + \alpha_6^H r_\perp^6\right) e^{-\nu_H r_\perp} \left(r_\perp^2 s\right)^{\lambda_H}, \quad (55)$$

$$\alpha_2^H = 0.072, \ \alpha_6^H = 1.89, \ \nu_H = 3.27, \ \lambda_H = 0.44.$$

To take into account phenomenologically the effects of gluon saturation we modified the exponential term in the formula (55) [26]. Namely, we assumed that $\nu_H$ slowly increases with a decrease of $x$. More exactly, we introduced the dependence of $\nu_H$ on $r_\perp^2 s$ rather than on $x$ having in mind that $r_\perp^2 \sim Q^{-2}$. We parametrized this dependence by the formula

$$\nu_H(r_\perp^2, s) = \frac{\nu_H}{\sqrt{1 - 0.7 e^{-\alpha\left(\frac{1}{r_\perp^2 s x_0}\right)^\lambda}}}, \quad (56)$$

$$\alpha = 10, \ \lambda = 0.288, \ x_0 = 3 \cdot 10^{-4}.$$

On Fig.1 we show the dependence of $\sigma_{q\bar{q}N}^{hard}$ on $r_\perp$ for several values of $r_\perp^2 s$. Since $r_\perp^2 s \sim x^{-1}$, for characteristic value of $r_\perp$, $r_\perp \sim 0.3$ fm and for $Q^2 \sim 0.4$ GeV$^2$, one can see from this figure that saturation effects, in our model, are essential at $x \leq 10^{-5}$.

On Fig.2 we show how the soft shadowing coefficient, $\alpha_{soft}$, is composed from contributions of vector mesons and high mass continuum. One can conclude, from this figure, that a contribution of the continuum to shadowing cannot be neglected even at very small values of $Q^2$.

On Fig.3 $Q^2$-dependencies of the soft shadowing coefficient are shown, for different values of $x$. One can see that the soft shadowing coefficient is small in a case of the pure $\rho$-dominance, if $Q^2 > M_{\rho_0}^2$. It is clearly visible also that all $Q^2$-dependencies are rather smooth.

On Fig.4 we show together the soft and hard coefficients, $\alpha_{soft}$ and $\alpha_{hard}$, and, on the same figure, the total coefficient $\alpha$ calculated using eq.(52), for two values of $Q^2$. One can see that a contribution of the hard shadowing correction $\delta\sigma_{\gamma^*A}^{hard}$ to $\alpha$ is negligibly small everywhere (and, correspondingly, the hard shadowing coefficient, $\alpha_{hard}$ is close to 1) but the hard cross section $\sigma_{\gamma^*N}^{hard}$ in the denominator of eq. (52) is essential, especially at large values of $Q^2$. Due to this cross section the behavior of the shadowing curve can become non-monotonous.

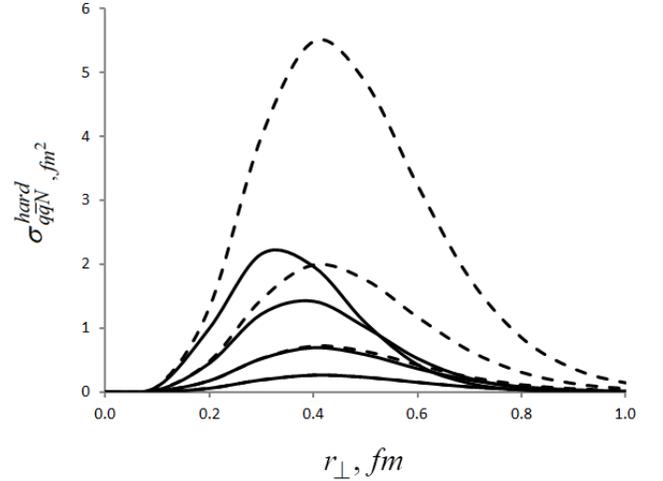

FIG. 1. The cross section of $q\bar{q} - N$ hard interaction as a function of $r_\perp$. From up to down: $(r_\perp s)^{-1} = 10^{-6}, 10^{-5}, 10^{-4}, 10^{-3}$. Dashed curves: eq. (55), solid curves: eq. (55) with modified $\nu_H$.

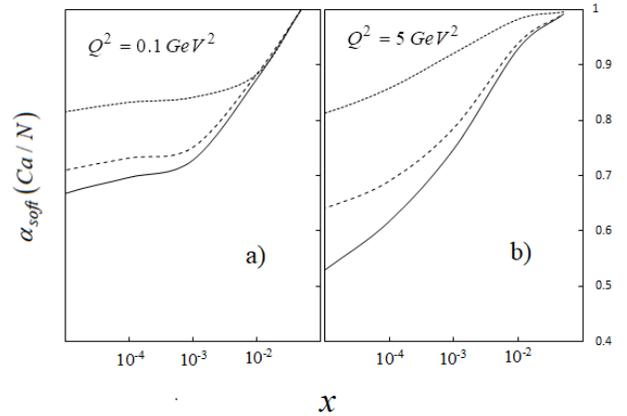

FIG. 2. a) The soft shadowing coefficient defined by eq. (25), for $^{40}$Ca as a function of $x$, for $Q^2 = 0.1$ GeV$^2$. Dotted curve: the contribution of $\rho_0$ only, dashed curve: $\rho_0$ +8 excited vector mesons, solid curve: the total sum (all separated vector mesons + continuum). b) the same as a) but for $Q^2 = 5$ GeV$^2$.

On Fig.5 the $x$-dependence of the total shadowing coefficient, $\alpha$, for two values of $Q^2$ is shown, for a case of xenon nucleus. For a comparison, the same calculation was performed without modification that takes into account the effects of gluon saturation (the corresponding result is shown by dashed curves). It is seen from the figure that at medium values of $Q^2$ the influence of saturation on a value of shadowing is noticeable beginning from $x \sim 10^{-4}$.

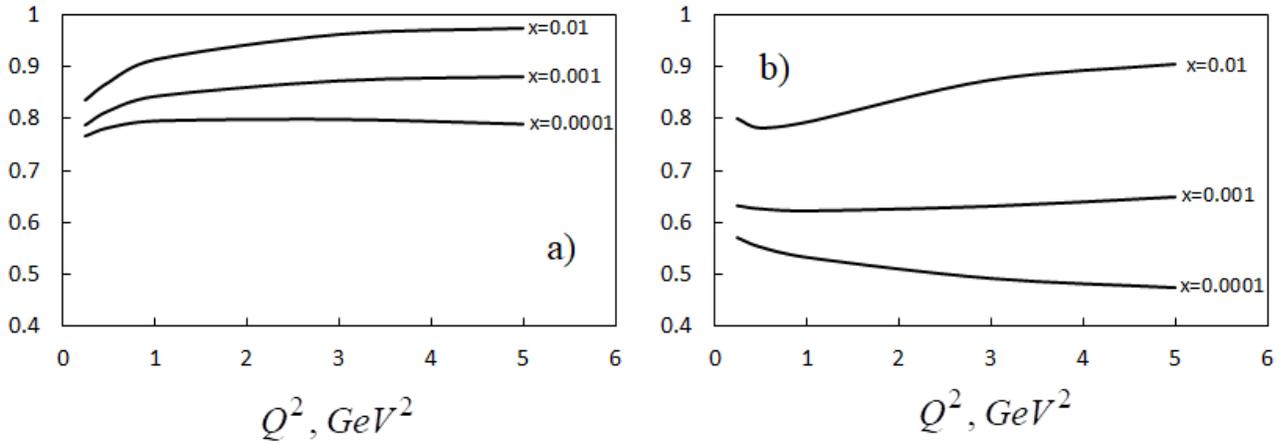

FIG. 3. The soft shadowing coefficient, $\alpha_{soft}$, for $^{132}$Xe as a function of $Q^2$, for three values of $x$. a) the contribution of $\rho_0$ only, b) the total soft coefficient.

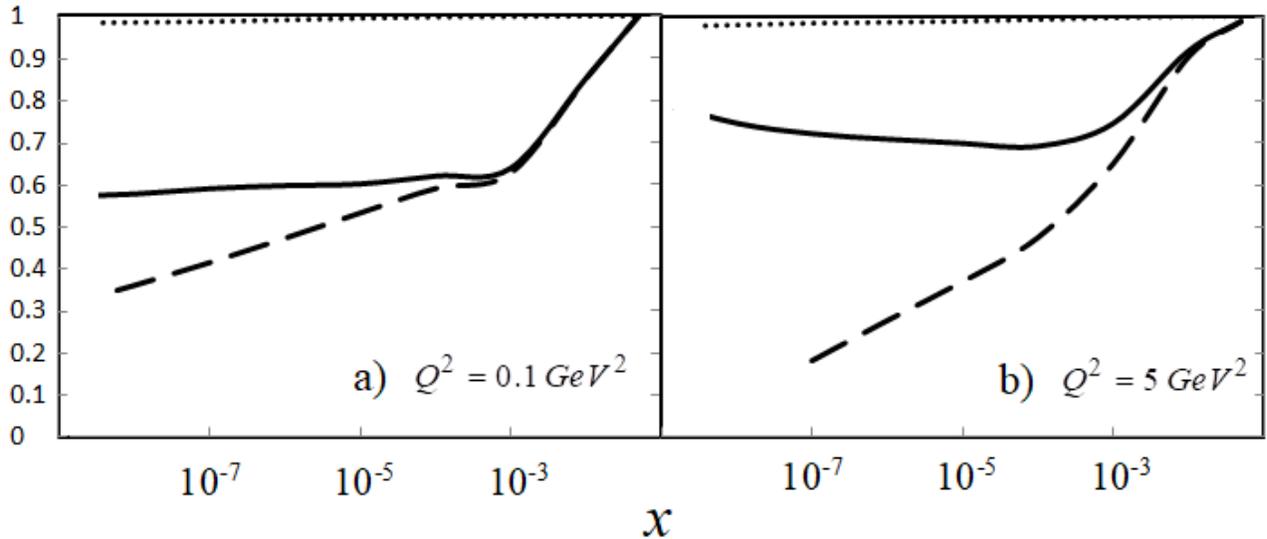

FIG. 4. Shadowing coefficients for $^{132}$Xe as a function of $x$, for fixed values of $Q^2$. a) dotted curve: the hard coefficient, eq. (51), dashed curve: the soft coefficient, eq. (41), solid curve: the total coefficient, eq. (52); b) the same as a) but for $Q^2 = 5\ GeV^2$.

EDGAR V. BUGAEV AND BORIS V. MANGAZEEV NUCLEAR SHADOWING IN DEEP INELASTIC…

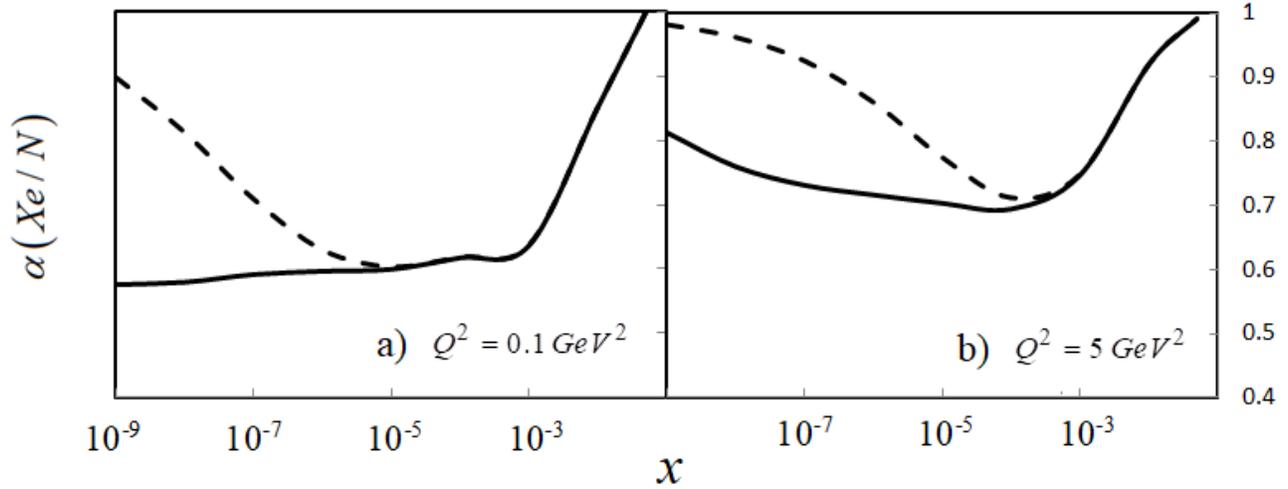

FIG. 5. The total shadowing coefficient, for $^{132}$Xe, as a function of $x$, for two values of $Q^2$. a) solid curve: our main result, dashed curve: the same, but without taking into account gluon saturation effects; b) the same as a) but for $Q^2 = 5\,GeV^2$.

On Fig.6 the A-dependence of $\alpha$ is shown for two values of $x$, for a characteristic value of $Q^2$. This figure shows the size of the predicted change of shadowing in a region of very small $x$ where there is no data.

On Fig.7 we show the comparison of results of our calculations of the shadowing coefficients with experimental data from E665 [46] and NMC [47] collaborations (the collection of experimental points is borrowed from the paper [48]). Different data points in all experiments correspond to different values of $Q^2$, from 0.5-0.6 GeV² at smallest values of $x$ up to a few GeV² at $x \sim 2\,10^{-2}$. The figure, largely, shows a quite reasonable agreement of our predictions with data, for all four nuclei.

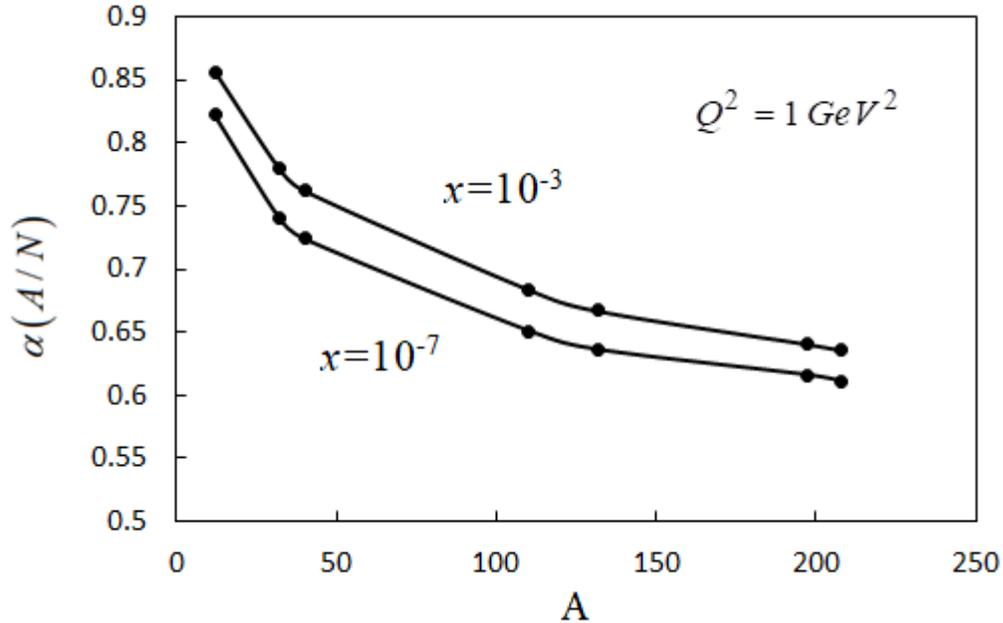

FIG. 6. A-dependence of the total shadowing coefficient, for fixed values of $x$ ($10^{-3}$, $10^{-7}$). For both curves the value of $Q^2$ is equal to 1 GeV². The separated points correspond to the nuclei $^{12}$C, $^{32}$S, $^{40}$Ca, $^{110}$Pd, $^{132}$Xe, $^{197}$Au, $^{208}$Pb.

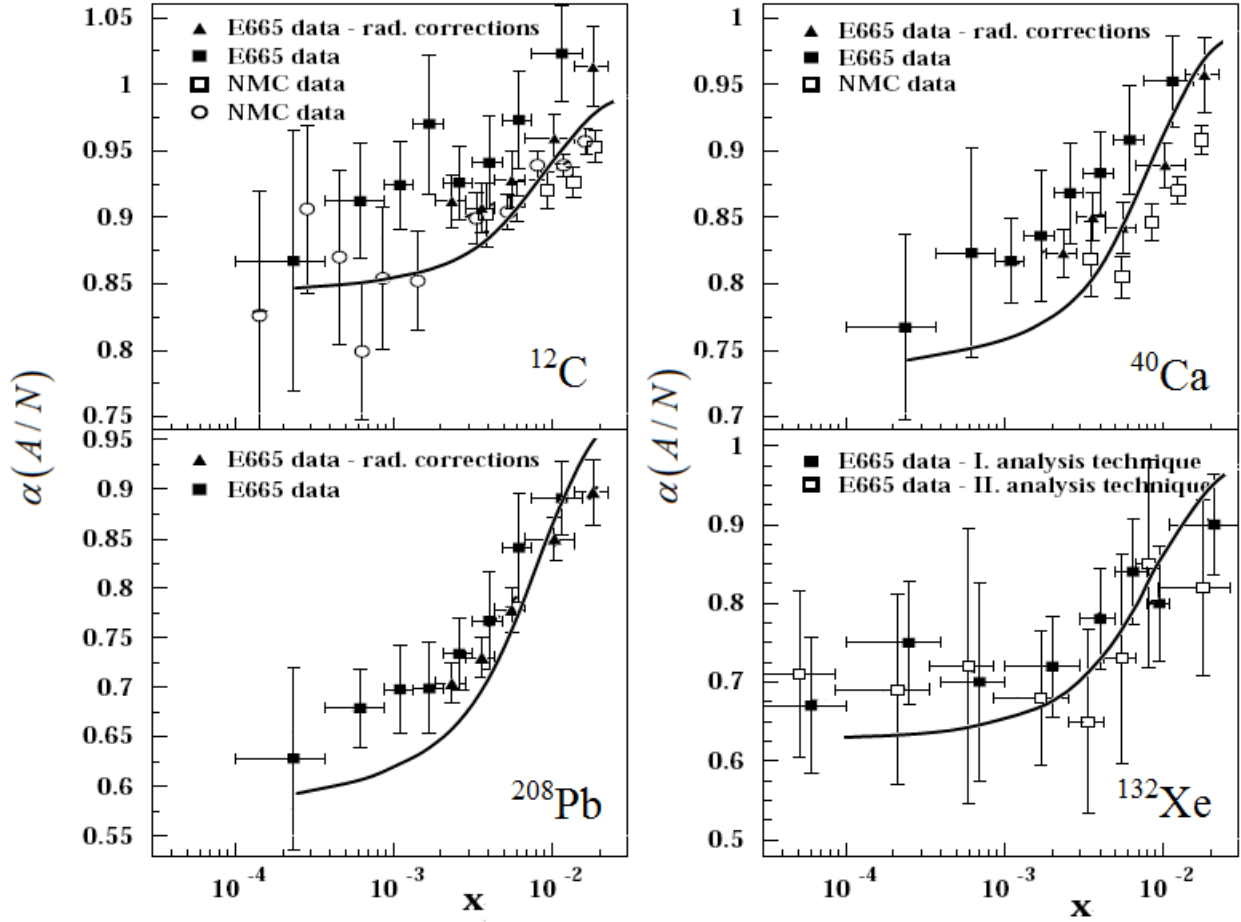

FIG. 7. Comparison of our predictions with experimental data from the E665 and the NMC collaborations.

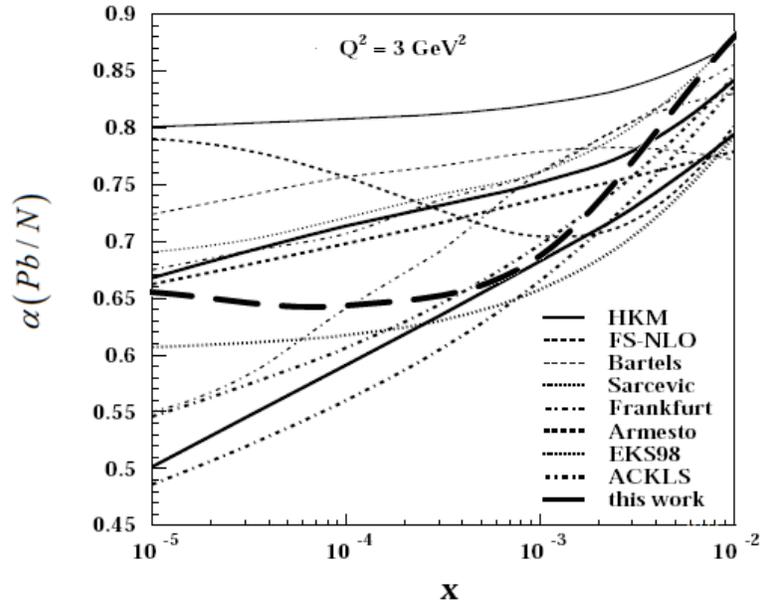

FIG. 8. Comparison of our prediction (thick dashed curve) for the total shadowing coefficient for $^{208}$Pb, at $Q^2 = 3$ GeV$^2$, with corresponding results of other works (the figure containing results of other authors is borrowed from [48]).

Finally, Fig. 8 shows a comparison of our predictions for nuclear shadowing for $^{208}$Pb, at $Q^2 = 3$ GeV$^2$, with the results of calculations using models of other authors. One can see that the disagreement between different predictions fastly grows with a decrease of $x$. This disagreement is rather large even at $x = (10^{-4}\text{-}10^{-5})$ where the gluon saturation effect which seems to be the main source of uncertainty is still relatively small.

## VI. CONCLUSIONS

In the present paper, for a description of the fluctuations of the virtual photon, i.e., intermediate states that interact strongly with the target nucleon, the hadronic (rather than quark-gluon) representation is used. Even when we consider the hard interactions of non-aligned $q\bar{q}$-pairs we use for a calculation of the corresponding cross section the phenomenological concepts of hadron dominance [49,50] and hard pomeron [39] rather than perturbative QCD directly (calculations of the hard part of $\sigma_{\gamma^* N}$ using the framework of perturbative QCD are carried out, e.g., in [52,53]). An use of the hadronic basis has, though, a large advantage: it allows to neglect non-diagonal transitions (such as $V(M) + N \to V(M') + N$) because, at high energies, the diffractive dissociation channels are suppressed (see, e.g., our work [26]).

Our calculation showed that the shadowing correction $\delta\sigma_{\gamma^* A}^{hard}$ is very small and can be completely neglected. This is the result of colour transparency, $\sigma_{q\bar{q}N}^{hard} \sim r_\perp^2$. As follows from Fig.1, a square of the hard cross section, $\left(\sigma_{q\bar{q}N}^{hard}\right)^2$, has a narrow peak at $r_\perp^{peak} \sim 0.3$ fm. At these $r_\perp$ a square of the photon wave function, $\left|\psi(r_\perp, Q^2)\right|^2$, is small because $\left(r_\perp^{peak}\right)^2 Q^2 \geq 1$ for characteristic values of $Q^2$, $Q^2 \sim 0.5$-1 GeV$^2$. Therefore, the integral over $r_\perp$ in eq. (46) is also small. Neglecting the correction $\delta\sigma_{\gamma^* A}^{hard}$ we are left with two sources of shadowing: i) coherent interactions (rescatterings) of discrete vector mesons and ii) coherent interactions of hadronic states of high-mass continuum. The corresponding shadowing corrections are $\delta\sigma_{\gamma^* A}^{(\rho, \rho'...)}$ and $\delta\sigma_{\gamma^* A}^{(cont)}$. Parameterizing the diffractive structure function $x_P F_2^{D(3)}$, in the quark-gluon representation, one has several terms [21],

$$x_P F_2^{D(3)} \simeq C_T F_{q\bar{q}}^T + C_L F_{q\bar{q}}^L + C_g F_{q\bar{q}g}^T, \quad (57)$$

corresponding to a production of $q\bar{q}$-pairs and $q\bar{q}g$ systems by the virtual photons. We assumed, in eq. (39), that our high-mass continuum arises from ($q\bar{q}g$ + high-mass $q\bar{q}$) - part of $x_P F_2^{D(3)}$ while $q\bar{q}$-pairs with low and intermediate masses form discrete vector mesons of the $\rho$-family.

Hard component of the hadronic fluctuations doesn't contribute to shadowing but it affects a behavior of the shadowing curve ($\alpha(x)$ at fixed $Q^2$) because the relative contribution of the hard component to $F_{2N}$ grows with a decrease of $x$. This leads to a change in the slope of the shadowing curves near $x \sim 10^{-3}$-$10^{-4}$, for all values of $Q^2$.

At smallest values of $x$ a modification of the formula for the hard cross section, $\sigma_{q\bar{q}N}^{hard}$, seems to be necessary, due to an influence of gluon saturation effects. The modification used in the present paper was suggested in [26]. The predictions of [26] for the nucleon structure function $F_{2N}$ at smallest $x$ and $Q^2$ obtained with an use of this modification are close to those of GBW model [53]. As shown in the paper (see Fig. 5) an accounting for the gluon saturation effects is very essential for predictions of shadowing at $x \leq 10^{-4}$-$10^{-5}$.